\documentclass{ws-ijmpd}
\usepackage{clshan-math,clshan-dm-dd}
\newcommand{\imageswitch} [2] {#2}
\def \lsim {\:\raisebox{-0.7ex}{$\stackrel{\textstyle<}{\sim}$}\:}
\def \gsim {\:\raisebox{-0.7ex}{$\stackrel{\textstyle>}{\sim}$}\:}
\def \ignore#1 {}
\begin{document}
\markboth{Chung-Lin Shan}
 {Effects of residue backgrounds
  in direct detection experiments on
  identifying WIMP DM}
\catchline{}{}{}{}{}
%
%%%%%%%%%%%%%%%%%%%%%%%%%%%%%%%%%%%%%%%%%%%%%%%%%%%%%%%%%%%%%%%%%%%
%
\title{Effects of Residue Background Events
       in Direct Detection Experiments on
       Identifying WIMP Dark Matter%
%\footnote{For the title, try not to use more than 3 lines. 
%Typeset the title in 10~pt Times roman, uppercase and boldface.}
}
\author{\footnotesize Chung-Lin Shan%\footnote{Typeset names in
%8~pt roman, uppercase. Use the footnote to indicate the
%present or permanent address of the author.}
\\~\\}
\address{Department of Physics, National Cheng Kung University      \\
         No.~1, University Road,
         Tainan City 70101, Taiwan, R.O.C. \\~\\
         Physics Division, National Center for Theoretical Sciences \\
         No.~101, Sec.~2, Kuang-Fu Road,
         Hsinchu City 30013, Taiwan, R.O.C. \\~\\
%
%\footnote{State completely without abbreviations, the 
%affiliation and mailing address, including country. Typeset in 
%8~pt Times italic.}\\
         clshan@mail.ncku.edu.tw}
%
%\author{SECOND AUTHOR}
%
%\address{Group, Laboratory, Address\\
%City, State ZIP/Zone, Country\\
%second\_author@group.com}
%
\maketitle
\begin{history}
\received{Day Month Year}
\revised{Day Month Year}
\comby{Managing Editor}
\end{history}
\begin{abstract}
 We reexamine the model--independent data analysis methods
 for extracting properties of
 Weakly Interacting Massive Particles (WIMPs)
 by using data (measured recoil energies) from
 direct Dark Matter detection experiments directly and,
 as a more realistic study,
 consider a small fraction of residue background events,
 which pass all discrimination criteria and
 then mix with other real WIMP--induced signals
 in the analyzed data sets.
 In this talk,
 the effects of residue backgrounds
 on the determination of
 the WIMP mass
 as well as
 the spin--independent WIMP coupling on nucleons
 will be discussed.
\end{abstract}
 % effect.}
\section{Introduction}
 In our earlier work on the development of
 model--independent data analysis methods
 for extracting properties of
 Weakly Interacting Massive Particles (WIMPs)
 by using measured recoil energies
 from direct Dark Matter detection experiments directly%
 \cite{DMDDf1v,DMDDmchi,DMDDfp2,DMDDidentification-DARK2009},
 it was assumed that
 the analyzed data sets are background--free,
 i.e., all events are WIMP signals.
 Active background discrimination techniques
 should make this condition possible.
 For example,
 by using the ratio of the ionization to recoil energy,
 the so--called ``ionization yield'',
 combined with the ``phonon pulse timing parameter'',
 the CDMS-II collaboration claimed that
 electron recoil events can be rejected event--by--event
 with a misidentification fraction of $< 10^{-6}$.%
 \cite{Ahmed09b}
%
% On the other hand,
 The CRESST collaboration demonstrated also that
 the pulse shape discrimination (PSD) technique
 can distinguish WIMP--induced nuclear recoils
 from those induced by backgrounds
 by means of inserting a scintillating foil,
 which causes some additional scintillation light
 for events induced by $\alpha$-decay of $\rmXA{Po}{210}$
 and thus shifts the pulse shapes of these events
 faster than pulses induced by WIMP interactions in the crystal%
 \cite{CRESST-bg}.% Lang09a, Schmaler09},
\footnote{
 More details
 about background discrimination techniques and status
 see also e.g.,
 Refs.~\refcite{bg-papers}.
             % Aprile09a,
             % EDELWEISS-bg, % Broniatowski09,
             % Lang09b}). %, Armengaud09}.
}

 However,
 as the most important issue in all
 underground experiments,
 possible residue background events
 which pass all discrimination criteria and
 then mix with other real WIMP--induced events in our data sets
 should also be considered.
 Therefore,
 as a more realistic study,
 we take into account
 small fractions of residue background events
 mixed in experimental data sets
 and want to study
 how well the model--independent methods
 could extract the {\em input} WIMP properties
 by using these ``impure'' data sets
 and how ``dirty'' these data sets could be
 to be still useful.

 In this article,
 I focus on two properties of WIMP Dark Matter:
 the mass $\mchi$ and
 the spin--independent (SI) coupling on nucleons $f_{\rm p}$.
 More detailed discussions
 can be found in Refs.~\refcite{DMDDbg-mchi,DMDDbg-fp2}.
\section{Effects of residue background events}
 In our numerical simulations
 based on the Monte Carlo method,
 while the shifted Maxwellian velocity distribution%
 \cite{SUSYDM96,DMDDf1v}
 with the standard values of
 the Sun's orbital velocity
 and the Earth's velocity
 in the Galactic frame:
 $v_0 \simeq 220~{\rm km/s}$
 and
 $\ve = 1.05 \~ v_0$,
 and the Woods--Saxon form
 for the elastic nuclear form factor
 for the spin--independent WIMP--nucleus interaction%
 \cite{Engel91,SUSYDM96}
 have been used for generating WIMP--induced signals,
 a {\em target--dependent exponential} form
 for residue background events has been introduced%
 \cite{DMDDbg-mchi}:
\beq
   \aDd{R}{Q}_{\rm bg, ex}
 = \exp\abrac{-\frac{Q /{\rm keV}}{A^{0.6}}}
\~.
\label{eqn:dRdQ_bg_ex}
\eeq
 Here $Q$ is the recoil energy,
 $A$ is the atomic mass number of the target nucleus.
 The power index of $A$, 0.6, is an empirical constant,
 which has been chosen so that
 the exponential background spectrum is
 somehow {\em similar to},
 but still {\em different from}
 the expected recoil spectrum of the target nucleus
 (see Figs.~\ref{fig:dRdQ-bg-ex-Ge-000-100-20});
 otherwise,
 there is in practice no difference between
 the WIMP scattering and background spectra.
 Note that,
 the atomic mass number $A$
 has been used here
 just as the simplest, unique characteristic parameter
 in the % general
 analytic form (\ref{eqn:dRdQ_bg_ex})
 for defining the residue background spectrum
 for {\em different} target nuclei.
 It does {\em not} mean that
 the (superposition of the real) background spectra
 would depend simply/primarily on $A$ or
 on the mass of the target nucleus, $\mN$.

 Note also that,
 firstly,
 the exponential form (\ref{eqn:dRdQ_bg_ex})
 for residue background spectrum
 is rather naive;
 however,
 since we consider here
 {\em only a few (tens) residue} background events
 induced by perhaps {\em two or more} different sources,
 pass all discrimination criteria,
 and then mix with other WIMP--induced events
 in our data sets of ${\cal O}(100)$ {\em total} events,
 exact forms of different background spectra
 are actually not very important and
 this exponential spectrum
 should practically not be unrealistic.
 Secondly,
 our model--independent data analysis procedures
 requires only measured recoil energies
 from one or more experimental data sets
 with different target nuclei%
 \cite{DMDDf1v,DMDDmchi,DMDDfp2,DMDDidentification-DARK2009}.
 Hence,
 for applying these methods to future real direct detection data,
 the prior knowledge about (different) background source(s)
 is {\em not required at all}.

 Moreover,
 the maximal cut--off of the velocity distribution function
 has been set as $\vmax = 700$ km/s.
 The experimental threshold energy
 has been assumed to be negligible
 and the maximal cut--off energy
 is set as 100 keV.
 The background window
 (the possible energy range
  in which residue background events {\em can not be ignored},
  compared to some other ranges)
 has been assumed to be the same as
 the experimental possible energy range.
 Note here that
 the actual numbers of generated signal and background events
 in each simulated experiment
 are Poisson--distributed around their expectation values
 {\em independently},
 and the total event number in one experiment
 is then the sum of these two numbers;
 both generated signal and background events
 are treated as WIMP signals in our analyses.
 Additionally,
 we assumed that
 all experimental systematic uncertainties
 as well as the uncertainty on
 the measurement of the recoil energy
 could be ignored.
\subsection{On the measured recoil spectrum}
\begin{figure}[p!]
\begin{center}
\vspace{-0.5cm}
\imageswitch{}
{%\hspace*{-0.75cm}%
 \includegraphics[width=9.3cm]{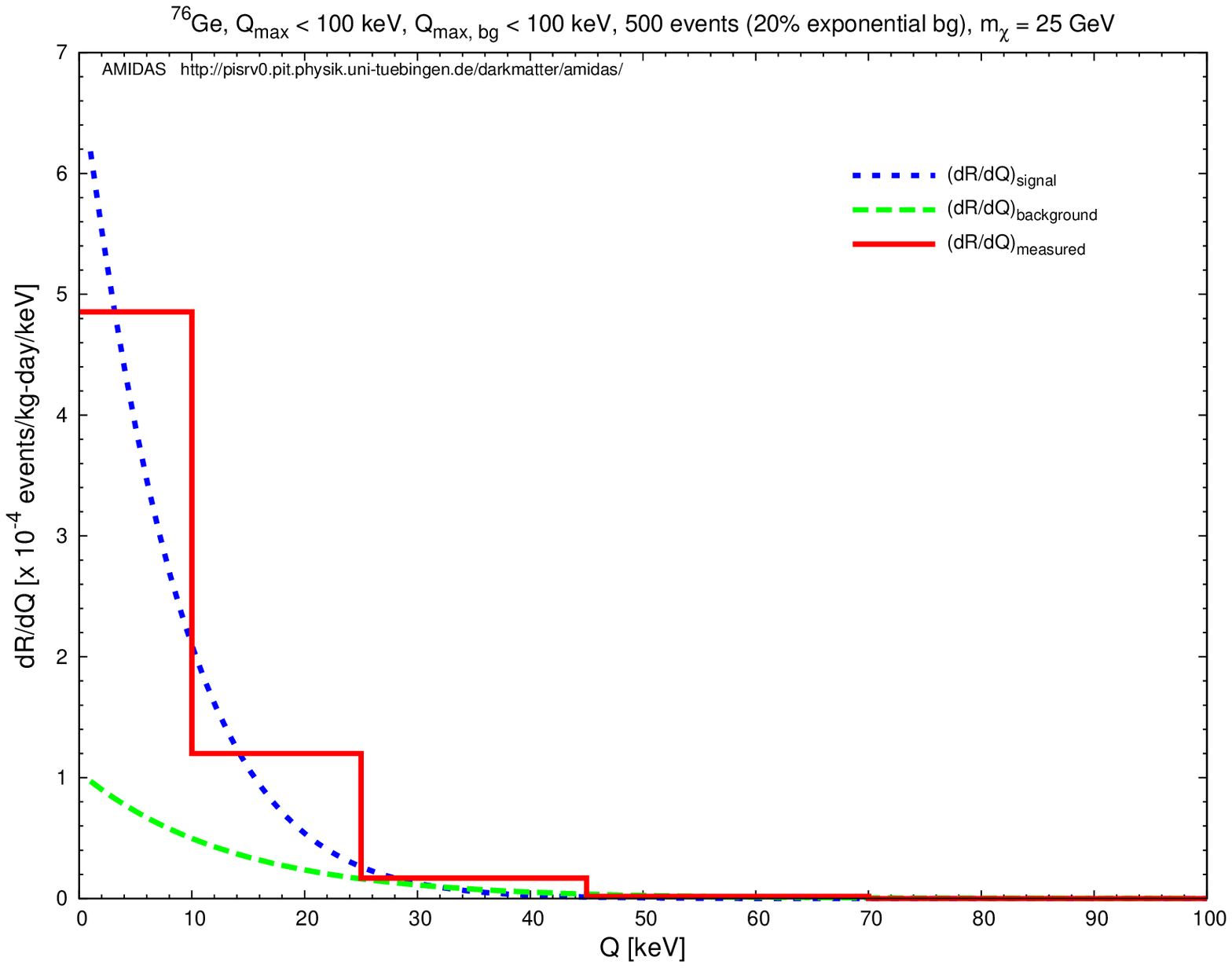} \\% \hspace*{-1.3cm} \\
% \vspace{0.2cm}%
% \hspace*{-0.75cm}%
 \includegraphics[width=9.3cm]{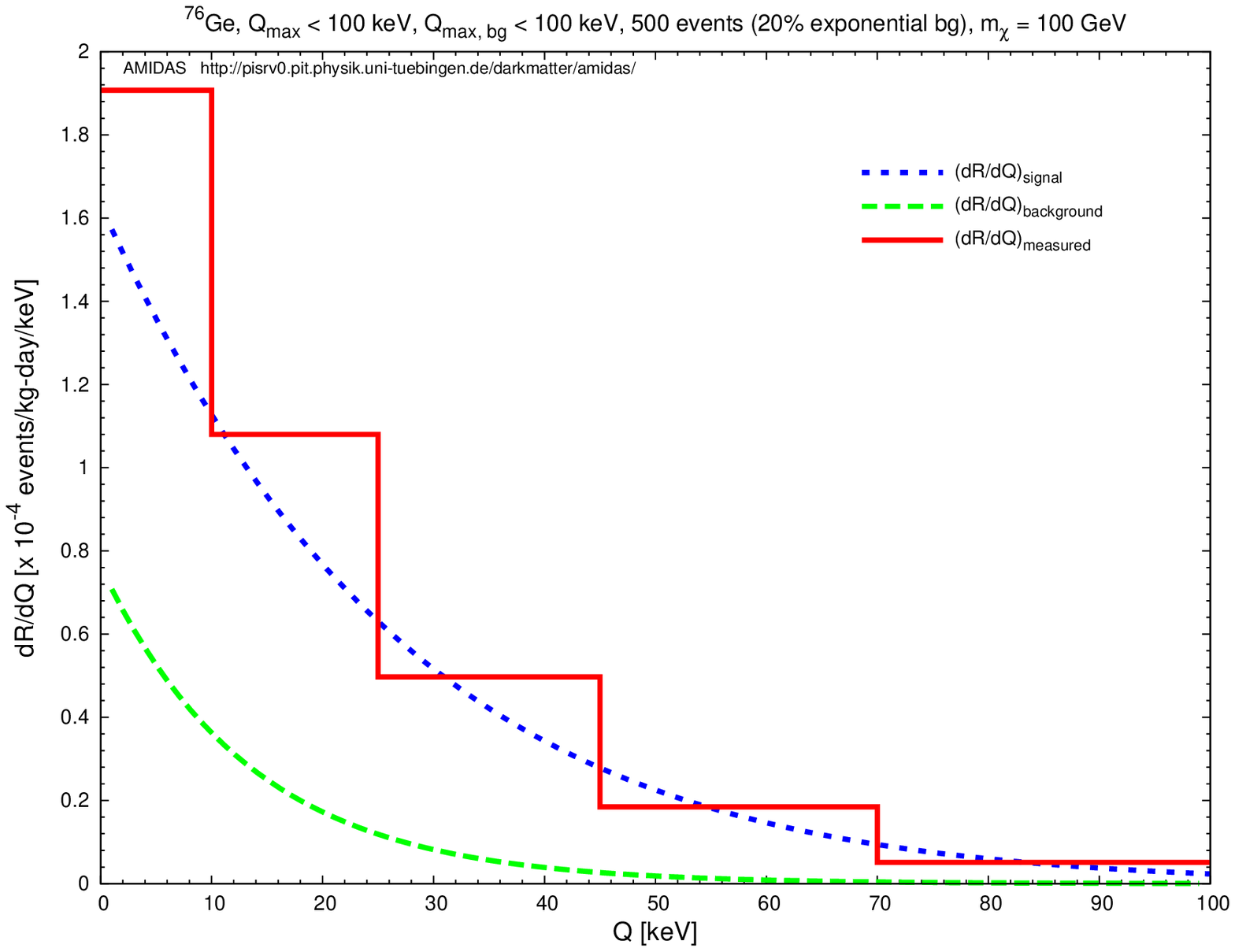} \\% \hspace*{-1.3cm} \\
% \vspace{0.2cm}%
% \hspace*{-0.75cm}%
 \includegraphics[width=9.3cm]{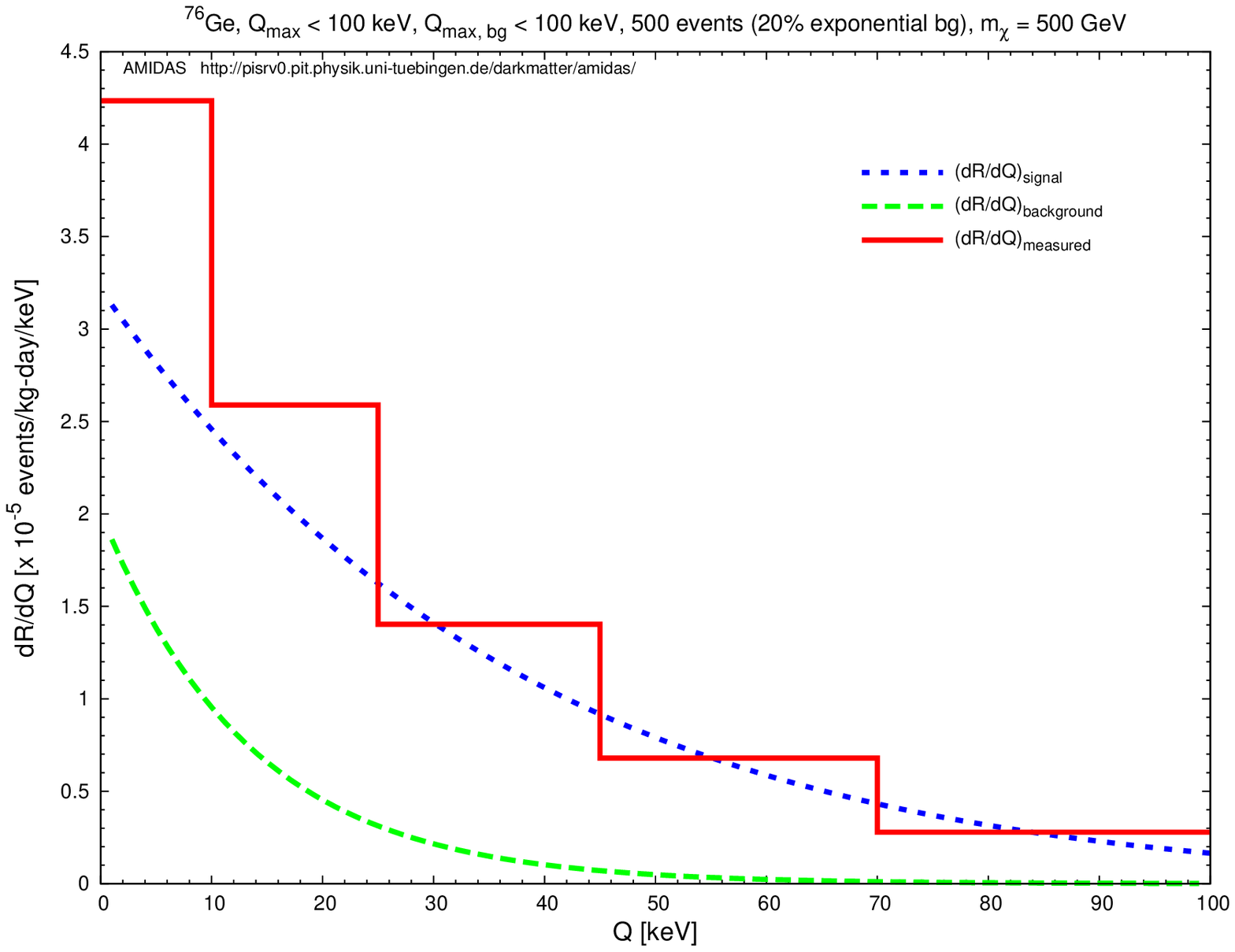} \\% \hspace*{-1.3cm} \\
}
\vspace{-0.5cm}
\end{center}
\caption{
 Measured energy spectra (solid red histograms)
 for a $\rmXA{Ge}{76}$ target
 with three different WIMP masses:
 25 (top), 100 (middle), and 500 (bottom) GeV.
 The dotted blue curves are
 the elastic WIMP--nucleus scattering spectra,
 whereas
 the dashed green curves are
 the exponential background spectra
 normalized to fit to the chosen background ratio,
 which has been set as 20\% here
 (plots from Ref.~8). % \cite{DMDDbg-mchi}).
}
\label{fig:dRdQ-bg-ex-Ge-000-100-20}
\end{figure}

 In Figs.~\ref{fig:dRdQ-bg-ex-Ge-000-100-20}
 I show measured energy spectra (solid red histograms)
 for a $\rmXA{Ge}{76}$ target
 with three different WIMP masses:
 25 (top), 100 (middle), and 500 (bottom) GeV.
 While the dotted blue curves show
 the elastic WIMP--nucleus scattering spectra,
 the dashed green curves indicate
 the exponential background spectrum
 given in Eq.~(\ref{eqn:dRdQ_bg_ex}),
 which have been normalized so that
 the ratios of the areas under these background spectra
 to those under the (dotted blue) WIMP scattering spectra
 are equal to the background--signal ratio
 in the whole data sets.
 5,000 experiments with 500 total events on average
 in each experiment have been simulated.

 It can be found here that,
 the shape of the WIMP scattering spectrum
 depends highly on the WIMP mass:
 for light WIMPs ($\mchi~\lsim~50$ GeV),
 the recoil spectra drop sharply with increasing recoil energies,
 while for heavy WIMPs ($\mchi~\gsim~100$ GeV),
 the spectra become flatter.
 In contrast,
 the exponential background spectra shown here
 depend only on the target mass
 and are rather {\em flatter}/{\em sharper}
 for {\em light}/{\em heavy} WIMP masses
 compared to the WIMP scattering spectra.
 This means that,
 once input WIMPs are {\em light}/{\em heavy},
 background events would contribute relatively more to
 {\em high}/{\em low} energy ranges,
 and, consequently,
 the measured energy spectra
 would mimic scattering spectra
 induced by {\em heavier}/{\em lighter} WIMPs.
 Moreover,
 for heavy WIMP masses,
 since background events would contribute relatively more to
 {\em low} energy ranges,
 the estimated value of the measured recoil spectrum
 at the experimental threshold energy %, $r(\Qmin)$,
% in Eq.~(\ref{eqn:rmin})
 could thus be (strongly) overestimated.
\subsection{On determining the WIMP mass}
\begin{figure}[t!]
\begin{center}
%\vspace{-0.75cm}
\imageswitch{}
{%\hspace*{-0.75cm}
 \includegraphics[width=9.3cm]{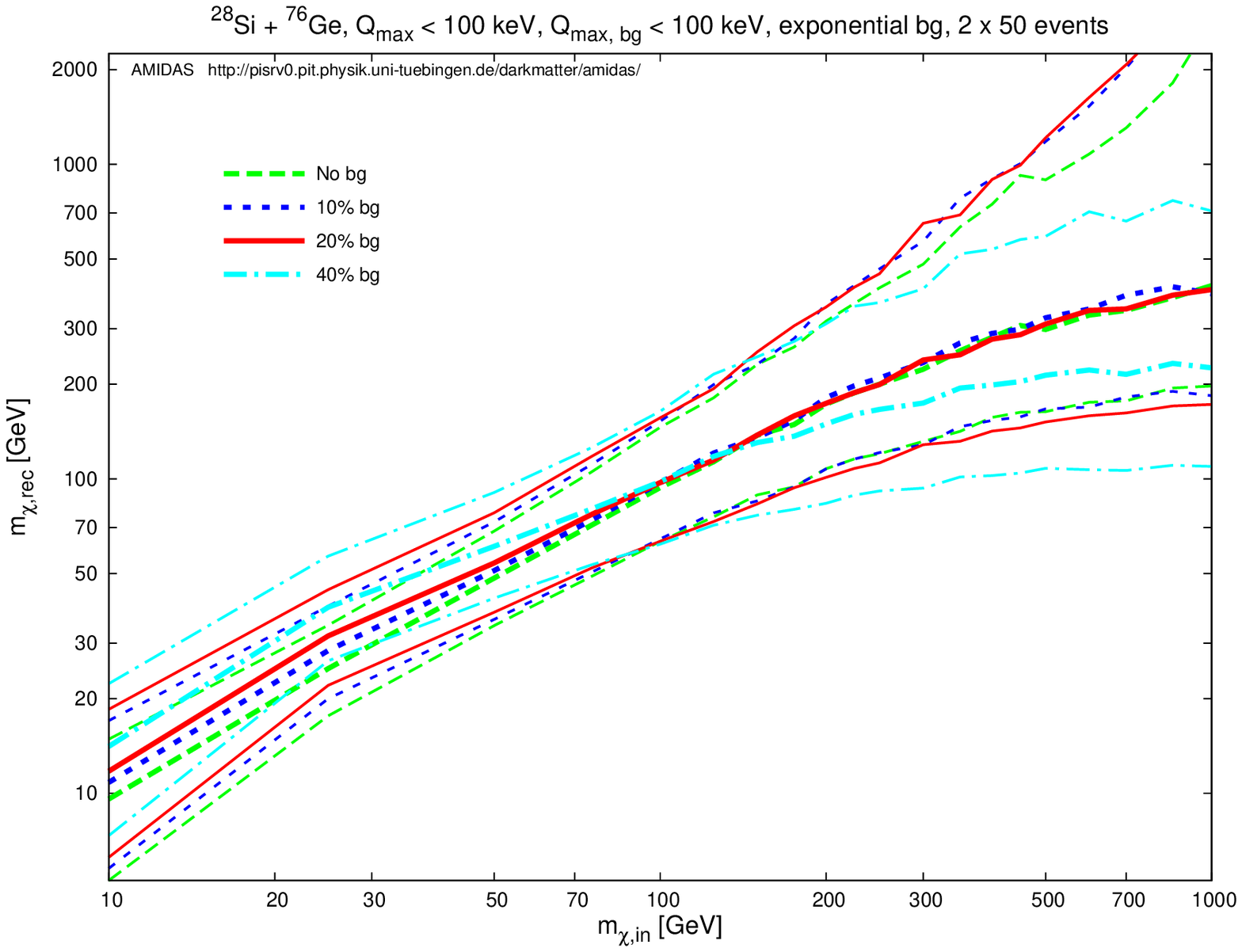}  \\
}
\vspace{-0.25cm}
\end{center}
\caption{
 The reconstructed WIMP masses
 as functions of the input WIMP mass.
 $\rmXA{Si}{28}$ and $\rmXA{Ge}{76}$
 have been chosen as two target nuclei.
 The background ratios shown here
 are no background (dashed green),
 10\% (long--dotted blue),
 20\% (solid red),
 and 40\% (dash--dotted cyan)
 background events in the analyzed data sets.
 Each experiment contains 50 total events
 on average.
 Other parameters are as
 in Figs.~1 % \ref{fig:dRdQ-bg-ex-Ge-000-100-20}
 (plot from Ref.~8). % \cite{DMDDbg-mchi}).
\vspace{-0.25cm}
}
\label{fig:mchi-SiGe-ex-000-100-050}
\end{figure}

 Fig.~\ref{fig:mchi-SiGe-ex-000-100-050}
 show the {\em median} values of
 the reconstructed WIMP mass
 and the lower and upper bounds of
 the 1$\sigma$ statistical uncertainty
 by means of the model--independent procedure introduced
 in Refs.~\refcite{DMDDmchi}
 with mixed data sets
 from WIMP--induced and background events
 as functions of the input WIMP mass.
 As in Refs.~\refcite{DMDDmchi},
 $\rmXA{Si}{28}$ and $\rmXA{Ge}{76}$
 have been chosen as two target nuclei.
 The background ratios shown here
 are no background (dashed green),
 10\% (long--dotted blue),
 20\% (solid red),
 and 40\% (dash--dotted cyan)
 background events in the analyzed data sets.
 2 $\times$ 5,000 experiments
 with 50 total events on average in each experiment
 have been simulated.

 It can be seen here clearly that,
 since
 for {\em light} WIMP masses ($\mchi~\lsim~100$ GeV),
 due to the relatively flatter background spectrum
 (compared to the scattering spectrum induced by WIMPs)
 or, in practice,
 some background sources in high energy ranges,
 the energy spectrum of all recorded events
 would mimic a scattering spectrum induced
 by WIMPs with a relatively {\em heavier} mass,
 the reconstructed WIMP masses
 as well as the statistical uncertainty intervals
 could be {\em overestimated}.
 In contrast,
 for {\em heavy} WIMP masses ($\mchi~\gsim~100$ GeV),
 due to the relatively sharper background spectrum
 or e.g., some electronic noise,
 relatively more background events
 contribute to low energy ranges,
 the energy spectrum of all recorded events
 would thus mimic a scattering spectrum induced
 by WIMPs with a relatively {\em lighter} mass.
 Hence,
 the reconstructed WIMP masses
 as well as the statistical uncertainty intervals
 could be {\em underestimated}.
 Nevertheless,
 Fig.~\ref{fig:mchi-SiGe-ex-000-100-050}
 shows that,
 with $\sim$ 20\% residue background events
 in the analyzed data sets
 of $\sim$ 50 total events,
 the 1$\sigma$ statistical uncertainty band
 can cover the true WIMP mass pretty well;
 if WIMPs are light ($\mchi~\lsim~200$ GeV),
 the maximal acceptable fraction of
 residue background events
 could even be as large as $\sim$ 40\%.

\subsection{On estimating the SI WIMP--nucleon coupling}

 In this section
 I show the {\em median} values of
 the reconstructed SI WIMP--nucleon coupling
 with its 1$\sigma$ statistical uncertainty
 by means of the model--independent method
 introduced in Refs.~\refcite{DMDDfp2}
 with mixed data sets.
 The SI WIMP--nucleon cross section for our simulations
 is set as $10^{-8}$ pb,
 the standard value for the local WIMP density,
 $\rho_0 = 0.3~{\rm GeV/cm^3}$,
% needed in Eq.~(\ref{eqn:fp2})
 has been used for both the simulations and data analyses.
 As in Refs.~\refcite{DMDDfp2},
 a $\rmXA{Ge}{76}$ nucleus has been chosen
 as our detector target for reconstructing $|f_{\rm p}|^2$;
 while a $\rmXA{Si}{28}$ target
 and a {\em second} $\rmXA{Ge}{76}$ target
 have been used for determining $\mchi$.
 The background ratios shown here
 are no background (dashed green),
%  5\% (dotted magenta lines),
 10\% (long--dotted blue),
 20\% (solid red),
 and 40\% (dash--dotted cyan)
 background events in the analyzed data set(s).
 (3 $\times$) 5,000 experiments
 with 50 total events on average in each experiment
 have been simulated.
\subsubsection{With a precisely known WIMP mass}
\begin{figure}[t!]
\begin{center}
%\vspace{-0.75cm}
\imageswitch{}
{%\hspace*{-0.75cm}
 \includegraphics[width=9.3cm]{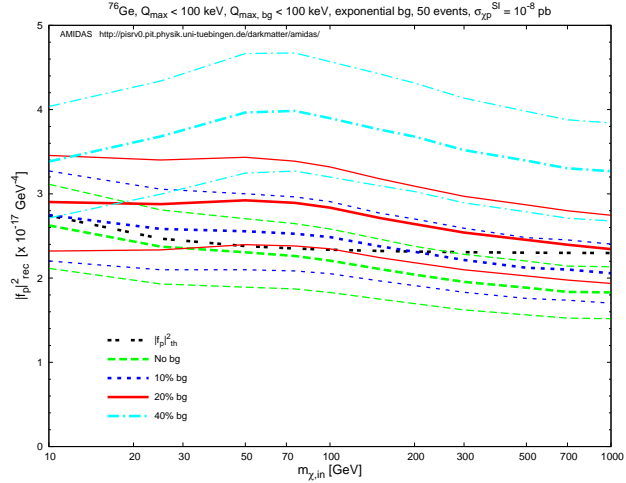}  \\
}
\vspace{-0.25cm}
\end{center}
\caption{
 The reconstructed SI WIMP--nucleon coupling
 as functions of the input WIMP mass.
 The double--dotted black curve
 is the theoretical value of $|f_{\rm p}|^2$
 corresponding to the fixed SI WIMP--nucleon cross section
 $\sigmapSI = 10^{-8}$ pb.
 The background ratios shown here
 are no background (dashed green),
%  5\% (dotted magenta lines),
 10\% (long--dotted blue),
 20\% (solid red),
 and 40\% (dash--dotted cyan)
 background events in the analyzed data set.
 Each experiment contains 50 total events
 on average.
 Other parameters are as in Figs.~1 % \ref{fig:dRdQ-bg-ex-Ge-000-100-20}
 (plot from Ref.~9). % \cite{DMDDbg-f1v}).
}
\label{fig:fp2-Ge-ex-000-100-050}
\end{figure}

 In Fig.~\ref{fig:fp2-Ge-ex-000-100-050}
 we first assume that
 the required WIMP mass
 for estimating $|f_{\rm p}|^2$
 has been known precisely
 from other (e.g., collider) experiments
 with an overall uncertainty of 5\% of
 the input (true) WIMP mass.
 It can be found
 in Fig.~\ref{fig:fp2-Ge-ex-000-100-050} that
 the {\em larger} the background ratio in the analyzed data set,
 the more strongly {\em overestimated}
 the reconstructed SI WIMP--nucleon coupling
 for {\em all} input WIMP masses.
 This can be understood
% from the expression (\ref{eqn:dRdQ}) for
% the different event rate $dR / dQ$.
 as follows.
 For a given WIMP mass and a specified target nucleus,
 the SI WIMP--nucleus cross section % in Eq.~(\ref{eqn:sigma0SI})
 is proportional to the total event number--to--exposure ratio.
 For a fixed number of total ``observed'' events,
 the {\em larger} the background ratio,
 or, equivalently,
 the {\em smaller} the number of real WIMP--induced events,
 the {\em smaller} the required exposure % $\calE = \calE_{\rm sg}$
 for accumulating the total observed events,
 and, therefore,
 the {\em larger} the estimated SI WIMP cross section/coupling.
 In other words,
 due to {\em extra unexpected} background events
 in our data set,
 one will use a {\em larger} number of total events
 to estimate the SI WIMP--nucleon coupling,
 and thus {\em overestimate} it.

 Moreover,
 it can also be seen in Fig.~\ref{fig:fp2-Ge-ex-000-100-050}
 that,
 in contrast to the reconstructed WIMP mass
 shown in Fig.~\ref{fig:mchi-SiGe-ex-000-100-050},
 where for an input WIMP mass of $\sim$ 100 GeV
 the effect of background events would be the smallest,
 the reconstructed SI WIMP--nucleon coupling
 would interestingly have the {\em largest deviation}
 for input WIMP masses between $\sim$ 50 GeV to 100 GeV,
 once the background ratio rises to $\gsim$ 20\%
 (the dash--dotted cyan curves).
\subsubsection{With a reconstructed WIMP mass}
\begin{figure}[t!]
\begin{center}
%\vspace{-0.75cm}
\imageswitch{}
{%\hspace*{-0.75cm}
 \includegraphics[width=9.3cm]{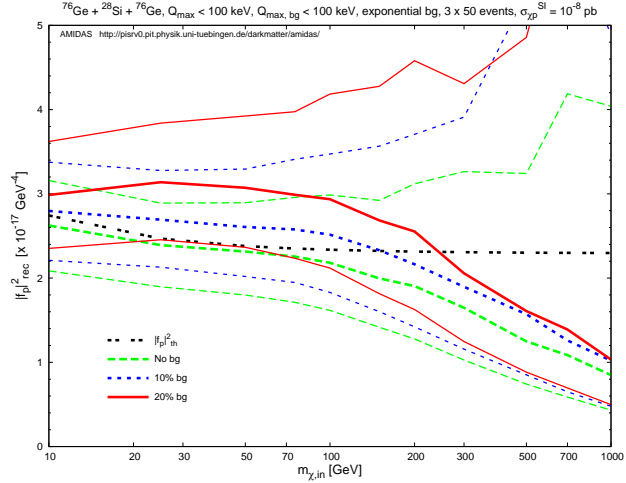}  \\
}
\vspace{-0.25cm}
\end{center}
\caption{
 As in Fig.~3, %\ref{fig:fp2-Ge-ex-000-100-050},
 except that
 the WIMP masses have been reconstructed
 by means of the procedure introduced
 in Refs.~2 % \cite{DMDDmchi}
 (plot from Ref.~9). % \cite{DMDDbg-f1v}).
\vspace{-0.2cm}
}
\label{fig:fp2-Ge-SiGe-ex-000-100-050}
\end{figure}

 In Fig.~\ref{fig:fp2-Ge-SiGe-ex-000-100-050}
 the required WIMP mass
 for estimating $|f_{\rm p}|^2$
 has been reconstructed
 with {\em other} direct detection experiments.
 As shown in Fig.~\ref{fig:mchi-SiGe-ex-000-100-050},
% and discussed in Sec.~3.2,
 due to the contribution from residue background events,
 if the input WIMP mass is {\em light}/{\em heavy},
 the reconstructed mass would be over-/underestimated.
 Hence,
 for input masses $\lsim$/$\gsim$ 150 GeV,
 the SI WIMP--nucleon coupling
 reconstructed by using three independent data sets
 would be {\em larger}/{\em smaller} than that
 reconstructed by using only one data set
 with extra information about the WIMP mass
 (cf.~Fig.~\ref{fig:fp2-Ge-ex-000-100-050}).
 In addition,
 the statistical uncertainty on the reconstructed SI WIMP coupling
 would also be (much) {\em larger}.
 However,
 Fig.~\ref{fig:fp2-Ge-SiGe-ex-000-100-050}
 indicates that
 one could in principle estimate the SI WIMP--nucleon coupling
 with an uncertainty of a factor $\lsim$ 2
 by using three independent data sets
 with maximal 20\% background events.
 For a WIMP mass of 100 GeV
 and a residue background ratio of 20\%,
% in entire experimental possible energy ranges,
 the deviation of the reconstructed SI WIMP coupling
 (with a reconstructed WIMP mass)
 would in principle be $\sim +13\%$
 with a statistical uncertainty
 of {$\sim^{+21\%}_{-14\%}$}
 ($\sim -3.3\%^{+18\%}_{-13\%}$
  for background--free data sets).
\section{Summary}
 In this article
 we reexamine the data analysis methods
 introduced in Refs.~\refcite{DMDDmchi,DMDDfp2}
 for determining the mass of Dark Matter particle
 and its spin--independent coupling on nucleons
 from measured recoil energies of direct detection experiments directly,
 by taking into account small fractions of residue background events,
 which pass all discrimination criteria and
 then mix with other real WIMP--induced events
 in the analyzed data sets.

 Our simulations show that,
 with a background ratio of $\sim$ 10\% -- 20\%
 in data sets of only $\sim$ 50 total events,
 while
 the 1$\sigma$ statistical uncertainty band
 of the reconstructed WIMP mass
 can cover the true value pretty well,
 especially for an input mass of $\sim$ 100 GeV,
 the reconstructed SI WIMP coupling on nucleons
 would be $\sim$ 10\% -- 15\% overestimated
 and the deviation would be the largest
 once the WIMP mass is between 50 and 100 GeV.
%
%\begin{figure}[pb]
%\centerline{\psfig{file=ijmpdf1.eps,width=4.7cm}}
%\vspace*{8pt}
%\caption{A schematic illustration of dissociative recombination. The
%direct mechanism, 4m$^2_\pi$ is initiated when the
%molecular ion S$_{\rm L}$ captures an electron with 
%kinetic energy. \label{f1}}
%\end{figure}
%
%\begin{table}[ph]
%\tbl{Comparison of acoustic for frequencies for piston-cylinder problem.}
%{\begin{tabular}{@{}cccc@{}} \toprule
%Piston mass & Analytical frequency & TRIA6-$S_1$ model &
%\% Error \\
%& (Rad/s) & (Rad/s) \\ \colrule
%1.0\hphantom{00} & \hphantom{0}281.0 & \hphantom{0}280.81 & 0.07 \\
%0.1\hphantom{00} & \hphantom{0}876.0 & \hphantom{0}875.74 & 0.03 \\
%0.01\hphantom{0} & 2441.0 & 2441.0\hphantom{0} & 0.0\hphantom{0} \\
%0.001 & 4130.0 & 4129.3\hphantom{0} & 0.16\\ \botrule
%\end{tabular} \label{ta1}}
%\end{table}
%
%
\section*{Acknowledgments}
 The author would like to thank
 the Physikalisches Institut der Universit\"at T\"ubingen
 for the technical support of the computational work
 demonstrated in this article.
 This work
 was partially supported by
 the National Science Council of R.O.C.~%
 under contracts no.~NSC-98-2811-M-006-044 and no.~NSC-99-2811-M-006-031
 as well as by
 the LHC Physics Focus Group,
 National Center of Theoretical Sciences, R.O.C..
%
%\appendix
%
%\section{Appendices}
%
%\section*{References}
%
%\begin{thebibliography}{000} %for 3 digits

%
%

\begin{thebibliography}{00}  %for 2 digits
%\begin{thebibliography}{0}    %for 1 digit
%
%% journal paper
%\bibitem{jpap} R. Loren and D. B. Benson, {\it J. Comput. 
%System Sci.} {\bf 27} (1983) 400.
%
%% collaboration
%\bibitem{colla} OPAL Collab. (G. Abbiendi {\it et al}.), 
%{\it Eur. J. Phys. C} {\bf 11} (1999) 217.
%
%% normal book (editors)
%\bibitem{edbk} R. Loren and D. B. Benson (eds.), {\it Introduction to 
%String Field Theory}, 2nd edn. (Springer-Verlag, New York, 1999).
%
%% review volume
%\bibitem{rvo} C. M. Wang, J. N. Reddy and K. H. Lee, New set of
%buckling parameters, in {\it Shear Deformable Beams}, ed.~T. Rex 
%(Elsevier, Oxford, 2000), p.~201.
%
%% book in a series
%\bibitem{seri} R. Loren, J. Li and D. B. Benson, Deterministic flow-chart 
%interpretations, in {\it Introduction to String Field Theory},  
%Advanced Series in Mathematical Physics, Vol.~3 (Springer-Verlag, New York, 1999), 
%p.~401.
%
%% proceedings
%\bibitem{pro} R. Loren, J. Li and D. B. Benson, Deterministic
%flow-chart interpretations, in {\it Proc. 3rd Int. Conf. 
%Entity-Relationship Approach}, eds. C. G. Davis and R. T. Yeh 
%(North-Holland, Amsterdam, 1983), p.~421.
%
%% to be published
%\bibitem{publ} R. Loren, J. Li and D. B. Benson, Deterministic
%flow-chart interpretations, to appear in {\it J. Comput. System Sci.} 
%
\bibitem{DMDDf1v}
 {M.~Drees and C.-L.~Shan,
%  {\it ``Reconstructing the Velocity Distribution of Weakly Interacting Massive Particles
%         from Direct Dark Matter Detection Data''},
  {\it J.~Cosmol.~Astropart.~Phys.}~{\bf 0706} (2007) 011.}
%  {\tt arXiv:astro-ph/0703651}.}
%
\bibitem{DMDDmchi}
%\bibitem{DMDDmchi-SUSY07}
 {M.~Drees and C.-L.~Shan,
% {C.-L.~Shan and M.~Drees,
%  {\it ``Determining the WIMP Mass from Direct Dark Matter Detection Data''},
 in {\it Proc.~15th Int.~Conf.~on
%  {\it proceedings of SUSY07},
%       the 15th International Conference on
       Supersymmetry and the Unification of Fundamental Interactions (SUSY07)},
  {\tt arXiv:0710.4296} (2007); % [hep-ph]} (2007);
%
%\bibitem{DMDDmchi}
% {M.~Drees and C.-L.~Shan,
%  {\it ``Model--Independent Determination of the WIMP Mass
%         from Direct Dark Matter Detection Data''},
  {\it J.~Cosmol.~Astropart.~Phys.}~{\bf 0806} (2008) 012.}
%  {\tt arXiv:0803.4477 [hep-ph]}.}
%
\bibitem{DMDDfp2}
%\bibitem{DMDDfp2-IDM2008}
 {M.~Drees and C.-L.~Shan,
%  {\it ``Constraining the Spin--Independent WIMP--Nucleon Coupling
%         from Direct Dark Matter Detection Data''},
%  {\it proceedings of % IDM 2008},
%       the 7th International Workshop on the Identification of Dark Matter (IDM 2008)},
  {\it PoS} {\bf IDM2008} (2008) 110;
%  {\tt arXiv:0809.2441 [hep-ph]}.}
%
%\bibitem{DMDDfp2}
% {M.~Drees and C.-L.~Shan,
  in preparation.}%}
%
\bibitem{DMDDidentification-DARK2009}
%\blue
 {M.~Drees and C.-L.~Shan,
%  {\it ``How Precisely Could We Identify WIMPs Model--Independently
%         with Direct Dark Matter Detection Experiments''},
 in {\it Proc.~7th Int.~Heidelberg Conf.~on
%  {\it proceedings of % DARK 2009},
%       the Seventh International Heidelberg Conference
       Dark Matter in Astro and Particle Physics (DARK 2009)},
  {\tt arXiv:0903.3300} (2009).} % [hep-ph]} (2009).}
%
%
% Background discrimination techniques and status
\bibitem{Ahmed09b}
%\blue
 {CDMS Collab.~(Z.~Ahmed {\it et al.}),
%  {\it ``Results from the Final Exposure of the CDMS II Experiment''},
  {\it Science} {\bf 327} (2010) 1619.}
%  {\tt arXiv:0912.3592 [astro-ph.CO]}.}
%
\bibitem{CRESST-bg}
%\bibitem{Lang09a}
 {CRESST Collab.~(R.~F.~Lang {\it et al.}),
%  {\it ``Discrimination of Recoil Backgrounds
%         in Scintillating Calorimeters''},
  {\it Astropart.~Phys.}~{\bf 33} (2010) 60;}
%  {\tt arXiv:0903.4687 [astro-ph.IM]}.}
%
%\bibitem{Schmaler09}
 {CRESST Collab.~(J.~Schmaler {\it et al.}),
%  {\it ``Status of the CRESST Dark Matter Search''},
  {\it AIP Conf.~Proc.}~{\bf 1185} (2009) 631.}
%  {\tt arXiv:0912.3689 [astro-ph.IM]}.}
%
\bibitem{bg-papers}
%\bibitem{Aprile09a}
 {E.~Aprile and L. Baudis, for the XENON100 Collab.,
%  {\it ``Status and Sensitivity Projections
%         for the XENON100 Dark Matter Experiment''},
  {\it PoS} {\bf IDM2008} (2008) 018;}
%  {\tt arXiv:0902.4253 [astro-ph.IM]}.}
%
%\bibitem{EDELWEISS-bg}
%\bibitem{Broniatowski09}
 {EDELWEISS Collab.~(A.~Broniatowski {\it et al.}),
%  {\it ``A New High--Background--Rejection
%         Dark Matter Ge Cryogenic Detector''},
  {\it Phys.~Lett.}~{\bf B 681} (2009) 305;}
%  {\tt arXiv:0905.0753 [astro-ph.IM]}.}
%
%\bibitem{Armengaud09}
 {EDELWEISS Collab.~(E.~Armengaud {\it et al.}),
%  {\it ``First Results of the EDELWEISS-II WIMP Search
%         Using Ge Cryogenic Detectors with Interleaved Electrodes''},
  {\it Phys.~Lett.}~{\bf B 687} (2010) 294;}
%  {\tt arXiv:0912.0805 [astro-ph.CO]}.}
%
%\bibitem{Lang09b}
 {CRESST Collab.~(R.~F.~Lang {\it et al.}),
%  {\it ``Electron and Gamma Background in CRESST Detectors''},
  {\it Astropart.~Phys.}~{\bf 32} (2010) 318.}
%  {\tt arXiv:0905.4282 [astro-ph.IM]}.}
%
%
\bibitem{DMDDbg-mchi}
 {Y.-T.~Chou and C.-L.~Shan,
%  {\it ``Effects of Residue Background Events
%         in Direct Dark Matter Detection Experiments on
%         the Determination of the WIMP Mass''},
  {\it J.~Cosmol.~Astropart.~Phys.}~{\bf 1008} (2010) 014.}
%  {\tt arXiv:1003.5277 [hep-ph]}.}
%
\bibitem{DMDDbg-fp2}
 {C.-L.~Shan,
  in preparation.}
%
%
\bibitem{SUSYDM96}
 {G.~Jungman, M.~Kamionkowski and K.~Griest,
%  {\it ``Supersymmetric Dark Matter''},
  {\it Phys.~Rep.}~{\bf 267} (1996) 195.}
%  {\tt arXiv:hep-ph/9506380}.}
%
% Woods-Saxon form factor
\bibitem{Engel91}
 {J.~Engel,
%  {\it ``Nuclear Form--Factors for the Scattering of
%         Weakly Interacting Massive Particles''},
  {\it Phys.~Lett.}~{\bf B 264} (1991) 114.}
%
\end{thebibliography}
\end{document}